\documentclass [12pt]{amsart}
\usepackage{amsmath,amssymb,dsfont}

\begin{document}
\title[Heat Polynoms, Umbral Correspondence and Burgers Eqs]{\bf Heat Polynomials, Umbral Correspondence and Burgers Equations.}


\author{G. Dattoli}
\address{G. Dattoli - ENEA CRE Frascati \\ via Enrico Fermi 45 00044 Rome Italy.\\ 
E-mail: \underline {dattoli@frascati.enea.it}}
\author{D. Levi}
\address{D. Levi - Dipartimento di Ingegneria Elettronica \\
Universit\`a degli Studi Roma Tre and Sezione INFN, Roma Tre \\
Via della Vasca Navale 84,00142 Roma, ITALY\\
E-mail: \underline {levi@fis.uniroma3.it}}

\bigskip






\begin{abstract}


We show that the umbral correspondence between differential equations can be 
achieved by means of a suitable transformation preserving the algebraic 
structure of the problems. We present the general properties of these 
transformations, derive explicit examples and discuss them in the case of 
the App\`{e}l and Sheffer polynomial families. We apply these 
transformations to non-linear equations, and discuss how the relevant 
solutions should be interpreted. 

\end{abstract}

\maketitle
\section{Introduction}


In this article we show the importance of the umbral calculus, together with 
the monomiality principle, for solving differential equations.

Umbral calculus \cite{1} provides us with a simple method to solve a large class 
of linear equations that are the umbral images of well known solvable 
differential equations. In this framework, we present some new umbral 
variables and use the umbral correspondence to solve modifications of the 
heat equation in terms of the solutions of the ordinary heat problem.

We consider the correspondence between the umbral methods and the more 
recently introduced principle of monomiality \cite{2}. According to the 
monomiality principle, classes of polynomials can be treated as ordinary 
monomials in suitable coordinates. These coordinates are obtained from a 
couple of suitable operators spanning a Weyl algebra and playing the role of 
derivative and multiplicative operators.

We represent explicitly the umbral correspondence by a transformation given 
in term of an operator that preserves the algebraic structure of the 
differential problems under study.
We consider the case of nonlinear systems and comment on the application of umbral and operational 
techniques in the  linearizable case. 

In Section 2 we introduce the 
umbral correspondence between different coordinate and ``differential'' 
operators by a transformation and apply it to the case of the heat equation 
and its modifications.
Section 3 is devoted to the discussion of the nonlinear Burgers equation and 
its modification when expressed in terms of different coordinates and 
``differential'' operators related by the umbral correspondence, while 
Section 4 is devoted to present some concluding remarks, including some 
considerations on a more general view on time ordering problems.

\bigskip

\section{ Umbral Correspondence and Heat Polynomials.}

According to ref. \cite{1} we can define the umbral image of a given differential 
equation
\begin{equation}
\label{eq1}
F(x,u(x),\partial _{x} u(x),\ldots ) = 0{\rm ,}
\end{equation}
\noindent
by the operator equation

\begin{equation}
\label{eq2}
F(\hat {q},u(\hat {q}),\hat {p}u(\hat {q}),\ldots ) = 0{\rm .}
\end{equation}
In eq. (\ref{eq2}) $\hat q, \, \hat p$
are some multiplicative and derivative operators such that 
\begin{equation}
\label{eq3}
{\left[ {\hat {p},\hat {q}} \right]} = \hat {1}{\rm ,}
\end{equation}
and $\hat {1}$ is a ``ground state'' on which the operators $\hat q, \, \hat p$
are acting\footnote{ Our operator algebra acts on this constant to generate 
its own irreducible representation, as it happens in the Lie algebraic 
treatment of special functions.}, usually a constant, which can be set in 
the whole generality equal to one. From eq. (\ref{eq3}) we get that formally we can 
always write:
\begin{equation}
\label{eq4}
\hat {p} = {\frac{{\partial} }{{\partial \hat {q}}}}{\rm .}
\end{equation}
The operator $\hat q$ will be referred as a multiplicative operator as, when acting on a basic polynomial of 
order n, P$_{{\rm n}}$($\hat {q}$), we have:
\begin{equation}
\label{eq5}
\hat {q}P_{n} (\hat {q}) = P_{n + 1} (\hat {q}){\rm .}
\end{equation}
The derivative operator $\hat p$ is an operator such that, when acting on a basic 
polynomial of order n, $P_{n} (\hat {q})$, we get:
\begin{equation}
\label{eq6}
\hat {p}P_{n} (\hat {q}) = nP_{n - 1} (\hat {q}){\rm .}
\end{equation}
From an abstract point of view, the two problems (\ref{eq1}) and (\ref{eq2}) are equivalent 
under the umbral correspondence, even though the operators, playing the role 
of multiplication and derivative, can be realized in totally different ways.

We can construct, starting from the operator equation (\ref{eq2}) a scalar equation 
by projecting it onto the ``ground state'' $\hat {1}$:
\begin{equation}
\label{eq7}
F(\hat {q},u(\hat {q}),\hat {p}u(\hat {q}),\ldots )\hat {1} = 0{\rm .}
\end{equation}
If eq. (\ref{eq1}) is an equation for $u(x)$ and $u(x)$ is an entire solution, than 
$u(\hat {q})$ will be the corresponding operator solution of eq. (\ref{eq2}). If eq. 
(\ref{eq1}) is a linear equation than $u(\hat {q})\hat {1}$ will be the solution of 
eq. (\ref{eq7}).

We can, in principle, always construct a map $\hat T$ which transform one set of 
coordinates and the corresponding ``differential'' operator into the other. 
If such a transformation $\hat {T}$ exists, it should have the following 
properties:
\begin{equation}
\label{eq8}
\begin{array}{l}
 \hat {T}\hat {T}^{ - 1} = \hat {1}, \qquad   \hat {T}x\hat {T}^{ - 1} = \hat {q}, \\ 
 \hat {T}\partial _{x} \hat {T}^{ - 1} = \hat {p},  \hat {T}\hat {1} = \hat {1}.  \\ 
 \end{array}
\end{equation}
From eqs. (\ref{eq8}) it follows that 
$
[\hat {p},\hat {q}] = \hat {T}[\partial _{x} ,x]\hat {T}^{ - 1}{\rm .}
$
Thus constant commutation relations are preserved. So if there exists an 
umbral map $\hat {T}$, for any choice of coordinates $\hat {q}$ and 
``differential'' operators $\hat {p}$, the commutation relation (\ref{eq3}) is 
satisfied. Moreover for any entire function f(x) we have:
\begin{equation}
\label{5b}
 \hat {T}f(x)\hat {T}^{ - 1} = f(\hat {q})^{}{\rm .}\end{equation}

We give here  an example of the construction of such a transformation. Let us consider the 
following $\hat {T}$ operator: 
\begin{equation}
\label{eq9}
\hat {T}_{y} = e^{y\partial^2 _{x}} 
\end{equation}
and with its inverse $\hat {T}_{y}^{ - 1} = e^{ - y\partial^2 _{x}} $, where by $\partial_x^2$ we mean $\partial_x \, \partial_x$. This 
transformation leaves the derivative operator unchanged, as
\begin{equation}
\label{eq10}
[\hat {T}_{y} ,\partial _{x} ] = 0{\rm ,}
\end{equation}
but it modifies the $x$ coordinate. From eq. (\ref{eq8}) the new coordinates and 
derivative operators are:
\begin{equation}
\label{eq11}
 \hat {q} = x + 2y\partial _{x} , \qquad
 \hat {p} = \partial _{x}.  
\end{equation}
It is immediate to prove that eqs. (\ref{eq5}, \ref{eq6}) are satisfied, together with any 
of their combinations. For example, we have: 
\begin{equation}
\label{eq12}
x\partial _{x} (x^{n}) = nx^{n} \quad {\rm ,}
\end{equation}
and, by the umbral correspondence, we will also have:
\begin{equation}
\label{eq13}
\hat {q}\hat {p}(\hat {q}^{n}) = n\hat {q}^{n}{\rm .}
\end{equation}

It is interesting to note that, according to the modified Burchnall identity 
\cite{2}, once $\hat {q}^{n}$ acts on unity we get:
\begin{equation}
\label{eq14}
\begin{array}{l}
\hat {q}^{n}\hat {1} = (x + 2y\partial _{x} )^{n}1 = {\sum\limits_{s = 
0}^{n} {\left( {{\begin{array}{*{20}c}
 {n} \hfill \\
 {s} \hfill \\
\end{array}} } \right)H_{n - s}} } (x,y)(2y\partial _{x} )^{s}1 = \\ \nonumber =
n!{\sum\limits_{r = 0}^{{\left[ {{\frac{{n}}{{2}}}} \right]}} {{\frac{{x^{n 
- 2r}y^{s}}}{{(n - 2s)!s!}}}}}  = H{}_{n}(x,y).
 \end{array}
\end{equation}
 $H_{n} (x,y)$ is a two variable Hermite polynomial of the Kamp\`{e} de 
F\`{e}ri\`{e}t type [3]\footnote{ The polynomals $H_{n} (x,y)$ are linked to 
the ordinary Laguerre polynomials by the identity $H_{n} (x,y) = 
i^{n}y^{{\frac{{n}}{{2}}}}H_{n} ({\frac{{ix}}{{2\sqrt {y}} }})$, where 
$H_{n} (x) = n!{\sum\limits_{r = 0}^{{\left[ {{\frac{{n}}{{2}}}} \right]}} 
{{\frac{{( - 1)^{r}(2x)^{n - 2r}}}{{(n - 2r)!r!}}}}} $.} . Eq. (\ref{eq14}) is just 
an example of the monomiality principle. Eq. (\ref{eq13}), which has as solution the 
monomials $\hat {q}^{n}$, can be rewritten as the Ordinary Differential 
Equation (ODE)
\begin{equation}
\label{eq15}
2yx\partial _{x} H_{n} (x,y) + \partial _{x,x} H_{n} (x,y) = nH_{n} 
(x,y){\rm .}
\end{equation}
Eq. (\ref{eq15}) is the ODE defining the family of basic polynomials (\ref{eq14}).

In conclusion the use of the transformation (\ref{eq9}) has shown the umbral 
equivalence between ordinary monomials $x^{n}$ and the Hermite polynomials 
$H_{n} (x,y)$, which, according to the language of Ref. \cite{2} are ``quasi monomials''.

Let us now consider the heat equation
\begin{equation}
\label{eq16}
\partial_t \Psi  = \partial_x^2 \Psi,
\end{equation}
for a function $\Psi = \Psi (x,t),$ with the initial condition 
\begin{equation}
\label{eq17}
\Psi \vert _{t = 0} = x^{n}{\rm .}
\end{equation}
The solution of eq. (\ref{eq16}, \ref{eq17}) can be formally written as:
\begin{equation}
\label{eq18}
\Psi = (e^{t\partial^2_{x}} x^{n})1{\rm .}
\end{equation}
Introducing the operator $\hat {T}_{t} = e^{t\partial^2_{x}} $, as in eq. 
(\ref{eq9}), eqs. (\ref{eq16}, \ref{eq17}) can be solved in terms of the two variable Hermite 
polynomials (\ref{eq14}):
\begin{equation}
\label{eq19}
\Psi = (\hat {T}_{t} x^{n})1 = \hat {q}^{n}\hat {1} = H_{n} (x,t) \quad {\rm 
.}
\end{equation}
For this reason, this family of polynomials is some time referred to as the 
heat polynomials (hp) \cite{4}.

In the previous example we have considered a particular initial condition 
consisting of an ordinary monomial. If we replace such an initial condition 
by a generic entire function 
\begin{equation}
\label{eq20}
\Psi \vert _{t = 0} = f(x){\rm ,}
\end{equation}
we can write the solution of the heat equation (\ref{eq16}, \ref{eq20}) as:
\begin{equation}
\label{eq21}
\Psi (x,t) = e^{t\partial^2_{x}} f(x) = f(\hat {q})\hat {1}{\rm .}
\end{equation}
In other words the umbrae $f(\hat {q})\hat {1}$ of the initial conditions 
$f(x)$ are solutions of the heat equations (\ref{eq16}, \ref{eq20}).

Let us consider now the action of the transformation (\ref{eq9}) with $y = \tau 
^{_{}} $ on the heat equation (\ref{eq16}, \ref{eq17}). According to the previous 
discussion we find
\begin{equation}
\label{eq22}
\begin{array}{l}
 \hat {T}_{\tau}  \partial_t \Psi = \hat {T}_{\tau}  \partial_x^2 \Psi  \Rightarrow \partial_t \Psi 
 (\hat {q},t)\hat {1} = \hat {p}^{2}\Psi (\hat {q},t)\hat {1}, \\ 
 \hat {T}_{\tau}  \Psi \vert _{t = 0} = \hat {T}_{\tau}  x^{n} \Rightarrow 
\Psi (\hat {q},0)\hat {1} = \hat {q}^{n}\hat {1}, \\ 
 \end{array}
\end{equation}
with $\hat {q} = x + 2\tau \partial _{x} $. The solution of eq. (\ref{eq22}) can 
now be written as:
\begin{equation}
\label{eq23}
\Psi \left( {\hat {q},t} \right)\hat {1} = \hat {T}_{t} \hat {q}^{n}1 = \hat 
{T}_{t} \hat {T}_{\tau}  x^{n}1 = H{}_{n}(x,t + \tau ){\rm .}
\end{equation}
The transformation (\ref{eq9}) transforms eq. (\ref{eq16}) into the following heat equation 
\begin{equation}
\label{eq24}
\Psi _{t} (x + 2\tau \partial _{x} ,t)1 = \partial_x^2 \Psi  (x + 2\tau \partial 
_{x} ,t)1,
\end{equation}
whose solution can be expressed in terms of the shifted in time two variable 
Hermite polynomials (\ref{eq23}).

Analogous results can be obtained using the transformations induced by the 
operators 
\begin{equation}
\label{eq25}
\hat {T}_{m,y} = e^{y\partial^m _{x}} ,m > 2{\rm .}
\end{equation}
In this case the corresponding coordinates and ``differential'' operators 
are:
\begin{equation}
\label{eq26}
 \hat {q} = x + my\partial _{x}^{m - 1}, \qquad
 \hat {p} = \partial _{x},
\end{equation}
and the corresponding basic polynomials, called higher order Hermite 
polynomials, are:
\begin{equation}
\label{eq27}
\hat {q}^{n}1 = H_{n}^{(m)} (x,y) = n!{\sum\limits_{r = 0}^{{\left[ 
{{\frac{{n}}{{m}}}} \right]}} {{\frac{{x^{n - mr}y^{r}}}{{(n - mr)!r!}}}} 
}{\rm .}
\end{equation}
So the solution of the higher order spatial derivative heat equation 
\begin{equation}
\label{eq28}
\begin{array}{l}
 \partial _{t} \Psi (x,t) = \partial^m_{x} \Psi (x,t),m > 2, \\ 
 \Psi (x,0) = x^{n}, \\ 
 \end{array}
\end{equation}
can be expressed in terms of the higher order Hermite polynomials (\ref{eq27}).

The similarity transformation (\ref{eq9}) transforms ordinary 
monomials into the two variable Hermite polynomials. This is not the only 
example. The App\`{e}l polynomials \cite{11} are characterized by an analogous 
property.

The App\`{e}l polynomials $a_{n} (x)$ are defined through the generating 
function \cite{11}
\begin{equation}
\label{eq29}
{\sum\limits_{n = 0}^{\infty}  {{\frac{{t^{n}}}{{n!}}}}} a_{n} (x) = 
A(t)e^{xt}{\rm ,}
\end{equation}
where $A(t)$is an undetermined function. Let us assume that there exists a 
domain for the variable $t$ where its Taylor expansion converges.
By the obvious identity
\begin{equation}
\label{eq30}
te^{xt} = \partial {}_{x}e^{xt}
\end{equation}
and by the assumption that $A(t)$is an entire function, we can rewrite eq. 
(\ref{eq29}) as:
\begin{equation}
\label{eq31}
{\sum\limits_{n = 0}^{\infty}  {{\frac{{t^{n}}}{{n!}}}}} a_{n} (x) = 
A(\partial _{x} )e^{tx}{\rm .}
\end{equation}
By expanding in power series the exponential in eq. (\ref{eq31}) and by equating 
the coefficients of the various powers of $t$, we end up with the following 
definition of the Appel polynomials:
\begin{equation}
\label{eq32}
a_{n} (x) = A(\partial _{x} )x^{n}{\rm .}
\end{equation}
The operator $A(\partial _{x} )$ will be referred to as the App\`{e}l 
operator. Let us assume that also its inverse $[A(\partial _{x} )]^{ - 1}$ 
is well defined so that
\begin{equation}
\label{eq33}
[A(\partial _{x} )]^{ - 1}A(\partial _{x} ) = \hat {1}{\rm .}
\end{equation}
When $A(\partial _{x} ) = \hat {T}_{t} $, we get the Hermite polynomials 
(\ref{eq19}) while, by choosing $A(\partial _{x} ) = {\frac{{1}}{{1 - \partial _{x} 
}}}$ we get the Truncated Exponential Polynomials (TEP) \cite{11}:
\begin{equation}
\label{27}
 \bar {e}_{n} (x) = {\frac{{1}}{{1 - \partial _{x}} }}x^{n},\quad \mbox{ i.e} \quad \bar {e}_{n} 
(x) = n!{\sum\limits_{r = 0}^{n} {{\frac{{x^{r}}}{{r!}}}}} .\end{equation}
We can introduce the Appel transformation $\hat {T}_{A} $=$A(\partial _{x} 
)$, as it satisfies all conditions (\ref{eq8}). Then the Appel coordinates and 
derivatives are\footnote{ In deriving eq. (\ref{eq34}) we have used the identity 
${\left[ {f(\partial _{x} ),x} \right]} = {f}'(\partial _{x} )$} :
\begin{equation}
\label{eq34}
\begin{array}{l}
 \hat {q} = A(\partial _{x} )x[A(\partial _{x} )]^{ - 1} = x + 
{\frac{{{A}'(\partial _{x} )}}{{A(\partial _{x} )}}}, \\ 
 \hat {p} = A(\partial _{x} )\partial _{x} [A(\partial _{x} )]^{ - 1} = 
\partial _{x} . \\ 
 \end{array}
\end{equation}
The application of $\hat {T}_{A} $ on eq. (\ref{eq12}) yields
\begin{equation}
\label{eq35}
(x + {\frac{{{A}'(\partial _{x} )}}{{A(\partial _{x} )}}})\partial _{x} 
a_{n} (x) = na_{n} (x){\rm .}
\end{equation}

In the case of the TEP, eq. (\ref{eq35}) reads:
\begin{equation}
\label{eq36}
x\partial^2 _{x} \bar {e}_{n} (x) - (x + n)\partial _{x} \bar {e}_{n} (x) + 
n\bar {e}_{n} (x) = 0{\rm .}
\end{equation}
As the derivative is invariant under an App\`{e}l transformation $\hat 
{T}_{A} $, the heat equation writes:
\begin{equation}
\label{eq37}
\partial _{t} \Psi (\hat {q},t) = \partial^2 _{x} \Psi (\hat {q},t){\rm .}
\end{equation}
If the initial condition of our problem is specified by
\begin{equation}
\label{eq38}
\Psi (\hat {q},t)\vert _{t = 0} = \hat {q}^{n}
\end{equation}
we can derive the correspondent associated heat polynomials as
\begin{equation}
\label{32a} H_{n} (\hat {q},t)1 = (e^{t\partial^2 _{x}} \hat {q}{}^{n})1 ={}_{A}H_{n} 
(x,t) = n!{\sum\limits_{r = 0}^{{\left[ {{\frac{{n}}{{2}}}} \right]}} 
{{\frac{{a_{n - 2r} (x)t^{r}}}{{(n - 2r)!r!}}}}} {\rm .}\end{equation} This family of 
polynomials satisfies the following recurrences
 \begin{equation}
\label{32b}\begin{array}{l}
 \partial _{x} {}_{A}H_{n} (x,t) = n{}_{A}H_{n - 1} (x,t), \\ 
 (x + {\frac{{A'(\partial _{x} )}}{{A(\partial _{x} )}}} + 2t\partial _{x} 
){}_{A}H_{n} (x,t) = {}_{A}H_{n + 1} (x,t). \\ 
 \end{array}\end{equation}

We have shown here a simple procedure to obtain the $\hat {T}$ operator 
which generate the umbral correspondence in the case of linear equations 
with constant coefficients. On the basis of such a correspondence we can 
define a wide class of Hermite type polynomials, playing the role of heat 
polynomials for various differential equations which are in umbral 
correspondence with the heat equation.

The situation is different if the equation of interest has $x$-dependent 
coefficients like, for example:
\begin{equation}
\label{eq39}
 \partial _{t} F(x,t) = xF(x,t) - \partial _{x} F(x,t), \quad
 F(x,0) = f(x). \\ 
\end{equation}
In this case the umbral counterpart writes:
\begin{equation}
\label{eq40}
 \partial _{t} F(\hat {q},t) = \hat {q}F(\hat {q},t) - \hat {p}F(\hat 
{q},t), \quad 
 F(\hat {q},0) = f(\hat {q}). \\ 
\end{equation}
Taking into account the 
Weyl decoupling identity: 
 \begin{equation}
\label{36}e^{\hat {A} + \hat {B}} = e^{ - {\frac{{k}}{{2}}}}e^{\hat {A}}e^{\hat 
{B}},\quad \mbox{if} \quad {\left[ {\hat {A},\hat {B}} \right]} = k \in C.\end{equation}
the general solution of eq. (\ref{eq40}) can be written, :
as\footnote{ Let us note that if ${\left[ {\hat {p},\hat {q}} \right]} = 1 
\to e^{\hat {p}t}f(\hat {q}) = f(\hat {q} + t)$.}  
\begin{equation}
\label{eq41}
F(\hat {q},t) = e^{(\hat {q} - \hat {p})t}f(\hat {q}) = e^{ - 
{\frac{{1}}{{2}}}t^{2}}e^{\hat {q}t}e^{ - \hat {p}t}f(\hat {q}) = e^{ - 
{\frac{{1}}{{2}}}t^{2}}e^{\hat {q}t}f(\hat {q} - t).
\end{equation}
Equation (\ref{eq39}) projected
 \begin{equation}
\nonumber
\partial _{t} F(\hat {q},t)\hat {1} = \hat {q}F(\hat {q},t)\hat {1} - \hat 
{p}F(\hat {q},t)\hat {1},\quad F(\hat {q},0)\hat {1} = f(\hat {q})\hat {1},
\end{equation}
will have the solution
\begin{equation}
\label{eq42}
F(\hat {q},t)1 = {\frac{{e^{ - {\frac{{1}}{{2}}}t^{2} - xt}}}{{1 - t}}}{\rm 
,}
\end{equation}
if $A(\partial _{x} ) = {\frac{{1}}{{1 - \partial _{x}} }}$ and $f(x) = 1$.

We will now show how the umbral transformation can be exploited to solve in 
a straightforward way modified heat equations. Let us consider the equation
\begin{equation}
\label{eq43}
\partial_t (t \partial_t \Psi  ) = \partial_x^2 \Psi {\rm .}
\end{equation}
Eq. (\ref{eq43}) can be thought a heat equation, in which the time derivative is 
replaced by 
\begin{equation}
\label{eq44}
\hat {p} = {}_{L}\hat {D}_{t} = \partial _{t} t\partial _{t} {\rm .}
\end{equation}
 $\hat {p}$ is some times referred to as the Laguerre derivative [5,6] and it 
satisfies the identity $\hat {p}^{n} = {}_{L}\hat {D}_{t}^{n} = \partial 
_{t}^{n} t^{n}\partial _{t}^{n} $.
We can associate to the Laguerre derivative a multiplicative operator 
\begin{equation}
\label{eq45}
\hat {q} = \partial _{t}^{ - 1} 
\end{equation}
such that\footnote{ We define $\partial _{t}^{ - 1} $ as the inverse of the 
derivative operator so that $\partial _{t}^{ - 1} \partial _{t} = \partial 
_{t} \partial _{t}^{ - 1} = 1$ \cite{2,12}. This assumption only holds if $\partial 
_{t}^{ - 1} $ is applied to entire functions.} 
\begin{equation}
\label{eq46}
[\hat {p},\hat {q}] = [\partial _{t} t\partial _{t} ,\partial _{t}^{ - 1} ] 
= 1{\rm .}
\end{equation}
The action of the operator $\partial _{t}^{ - n} $on a given function $f(t)$ 
can be written as a Cauchy repeated integral, namely
\begin{equation}
\label{eq47}
\partial _{t}^{ - n} f(t) = {\frac{{1}}{{\Gamma (n + 
1)}}}{\int\limits_{0}^{\infty}  {(t - \xi )^{n}}} f(\xi )d\xi {\rm .}
\end{equation}
From eq. (\ref{eq47}) follows immediately \cite{2,7} that
\begin{equation}
\label{eq48}
(\partial _{t}^{ - 1} )^{n}1 = {\frac{{t^{n}}}{{n!}}}{\rm .}
\end{equation}
In the transformed variables, eq. (\ref{eq43}) can be written as:
\begin{equation}
\label{eq49}
\hat {p}\Psi (x,\hat {q}) = \partial_x^2 \Psi  (x,\hat {q}){\rm .}
\end{equation}
The associated hp read:
\begin{equation}
\label{eq50}
{}_{2}H_{n} (x,t) = H_{n} (x,\hat {q})\hat {1} =  n!{\sum\limits_{r = 
0}^{{\left[ {{\frac{{n}}{{2}}}} \right]}} {{\frac{{x^{n - 2r}t^{r}}}{{(n - 
2r)1\left( {r!} \right)^{2}}}}}} {\rm .}
\end{equation}
The polynomial (\ref{eq50}) belongs to the family of hybrid polynomials. These polynomials are
situated in between the Hermite and Laguerre polynomials \cite{2,5}. By direct 
calculation one can prove that the polynomials (\ref{eq50}) solve eq. (\ref{eq43}).

In this example we have introduced the umbral correspondence between eq. 
(\ref{eq43}) and (\ref{eq49}) by defining the new coordinates and derivatives, ($\hat 
{q},\hat {p}$), given by eqs. (\ref{eq44}, \ref{eq45}). Let us look for the transformation 
$\hat {T}_{L} $ which provide the ``transition'' 
\begin{equation}
\label{eq51}
H_{n} (x,y) \to {}_{2}H_{n} (x,y){\rm .}
\end{equation}
Formally this transformation is obtained by requiring
\begin{equation}
\label{eq52}
\hat {T}_{L} t\hat {T}_{L}^{ - 1} = \partial _{t}^{ - 1} = \hat {q}{\rm .}
\end{equation}
The explicit form of $\hat {T}_{L} $ can be derived by noting that the 
correspondence between $t$ and $\hat {q}$ implies $t^{n} \to 
{\frac{{t^{n}}}{{n!}}}$. 

Given a function $f(t) = {\sum\limits_{n = 0}^{\infty}  {{\frac{{f_{n} 
}}{{n!}}}}} t^{n}$, taking into account eqs. (\ref{eq48}, \ref{5b}) we can write:
\begin{equation}
\label{eq53}
\hat {T}_{L} ^{ - 1}f(t) = {\int\limits_{0}^{\infty}  {e^{ - t\xi} }} f(x\xi 
)d\xi = {\sum\limits_{n = 0}^{\infty}  {f_{n}} } t^{n}{\rm .}
\end{equation}
Thus $\hat {T}_{L} $ can be viewed as the inverse of the Laplace transform 
(\ref{eq53}) (see also ref. \cite{8}).

In the previous paragraphs we have considered transformations associated 
with App\`{e}l polynomials which leave the derivative operator invariant. 
Here we will consider more general transformations, when also the derivative 
operator is changed. We will call one such transformation a Sheffer 
transformation.

The Sheffer polynomials $\sigma_{n} (x)$ are a natural extension of the App\`{e}l 
polynomials $a_{n} (x)$. They are generated by \cite{13}
\begin{equation}
\label{eq54}
{\sum\limits_{n = 0}^{\infty}  {{\frac{{t^{n}}}{{n!}}}}} \sigma_{n} (x) = 
A(t)\exp (xB(t)){\rm ,}
\end{equation}
where $B(t) $is, as $A(t)$, an entire function$.$ Following ref. \cite{6} we can prove the 
quasi-monomiality of $\sigma_{n} (x)$. If we take $A(t) = 1$ we can easily 
construct $\hat {p}$-operators which do not coincide with the derivative 
operator $\partial _{x} $ and the associated
 Sheffer polynomials $s_{n} (x)$ are given by
\begin{equation}
\label{eq55}
{\sum\limits_{n = 0}^{\infty}  {{\frac{{t^{n}}}{{n!}}}}} s_{n} (x) = \exp 
(xB(t)){\rm .}
\end{equation}
Multiplying both sides of eq. (\ref{eq55}) by the operator $B^{ - 1}(\partial _{x} 
)$, we find:
\begin{equation}
\label{eq56}
{\sum\limits_{n = 0}^{\infty}  {{\frac{{t^{n}}}{{n!}}}B^{ - 1}(\partial _{x} 
)}} s_{n} (x) = B^{ - 1}(\partial _{x} )\exp (xB(t)) = t\exp (xB(t)){\rm .}
\end{equation}
From eq. (\ref{eq56}) we deduce that the derivative operator for the Sheffer 
polynomials $s_{n} (x)$ is given by
\begin{equation}
\label{eq57}
\hat {p} = B^{ - 1}(\partial _{x} ){\rm .}
\end{equation}
The multiplicative operator can be obtained by taking the derivative with 
respect to $t $ of both sides of eq. (\ref{eq55}):
\begin{equation}
\nonumber
{\sum\limits_{n = 0}^{\infty}  {{\frac{{nt^{n - 1}}}{{n!}}}}} s_{n} (x) = 
x{B}'(t)\exp (xB(t)) = x{B}'(B^{ - 1}(\partial _{x} ))\exp (xB(t)){\rm .}
\end{equation}
Thus the multiplicative operator corresponding to $\hat {p}$ given by eq. 
(\ref{eq57}) takes the form
\begin{equation}
\label{53}
\hat {q} = x{B}'(\hat {p}).\end{equation} 
A typical example of coordinate and momenta operators associated to a 
Sheffer transform are
\begin{equation}
\label{eq59}
 \hat {p} = e^{\partial _{x}}  - 1, \qquad
 \hat {q} = xe^{ - \partial _{x}} , 
\end{equation}
when $B(\hat {p}) = \ln (\hat {p} + 1)$ and thus $B'(\hat {p}) = 
{\frac{{1}}{{\hat {p} + 1}}}$. As $e^{\partial_x} f(x) = f(x+1)$, the derivative $\hat p$ acts on a function $f(x)$ as a shift operator. The associated polynomials are the lower 
factorial specified by
\begin{equation}
\nonumber
s{}_{n}(x) = {\frac{{\Gamma (x + 1)}}{{\Gamma (x + 1 - n)}}} = (x){}_{n}{\rm 
.}
\end{equation}
The  heat equation, in which the spatial derivatives are substituted by $\hat p$ given by eq. (\ref{eq59}),  admits a solution given by the corresponding 
hp, namely
\begin{equation}
\label{eq61}
H_{n} (\hat {q},t)1 = n!{\sum\limits_{r = 0}^{{\left[ {{\frac{{n}}{{2}}}} 
\right]}} {{\frac{{(x)_{n - 2r} t^{r}}}{{(n - 2r)!r!}}}}} {\rm .}
\end{equation}

By a proper definition of the function $B(\hat {p})$ we can obtain any 
discrete representation of the derivative operator. For example, by choosing 
$B(\hat {p}) = \ln (\hat {p} + \sqrt {\hat {p}^{2} + 1} )$ we get $\hat {p} 
= {\frac{{e^{\partial _{x}}  - e^{ - \partial _{x}} }}{{2}}}$, $B'(\hat {p}) 
= {\frac{{1}}{{\sqrt {\hat {p}^{2} + 1}} }}$ and thus $\hat {q} = 
2x(e^{\partial _{x}}  + e^{ - \partial _{x}} )^{ - 1}$ \cite{1, 9}\footnote{ If $A(t) \ne 1$ the Sheffer polynomials 
can now be defined by the operational rule $\tilde {s}_{n} (x) = A(\hat 
{p})s_{n} (x)$. The transformation $\hat {T}_{S} $ (\ref{eq8}) is therefore 
specified by the following identities, which preserve the Weyl algebra 
structure: \par $\begin{array}{l}
 \hat {T}_{S} \partial _{x} \hat {T}_{S} ^{ - 1} = \hat {p} = B^{ - 
1}(\partial _{x} ), \\ 
 \hat {T}_{S} x\hat {T}_{S} ^{ - 1} = \hat {q} = x{B}'(\hat {p}) + 
{\frac{{{A}'(\hat {p})}}{{A(\hat {p})}}}. \\ 
 \end{array}$}.

We have considered so far the case in which the $\hat {T}$ transform has 
affected the spatial or the time components of the equation but not both.
Let us consider the heat equation
\begin{equation}
\label{eq62}
\begin{array}{l}
[ (e^{\partial _{t}}  - 1)\Psi] 1 = [(e^{\partial _{x}}  - 1)^{2}\Psi ]1, \\ 
 \Psi \vert _{t = 0} 1 = \left( {x} \right)_{n} , \\ 
 \end{array}
\end{equation}
whose solution will be generated by a double transformation.
We thus get:
\begin{equation}
\label{eq63}
\Psi (x,t) = n!{\sum\limits_{r = 0}^{{\left[ {{\frac{{n}}{{2}}}} \right]}} 
{{\frac{{(x)_{n - 2r} (t)_{r}} }{{(n - 2r)!r!}}}}} {\rm .}
\end{equation}
The solution of the heat equation
\begin{equation}
\label{eq64}
\ln (1 + \partial _{t} )\Psi 1 = (e^{\partial _{x}}  - 1)^{2}\Psi 1{\rm ,}
\end{equation}
can be obtained using transformations associated with Sheffer Bell-type 
polynomialsin the $t$--variable.  In this case we have:
\begin{equation}
\label{eq65}
 \hat {q} = t(1 + \partial _{t} ), \qquad
 \hat {q}^{n}1 = b_{n} (t) = {\sum\limits_{k = 1}^{n} {S_{2}} } (n,k)t^{k}. 
\end{equation}
In eq.(\ref{eq65}) $S_{2} (n,k)$ is a Stirling number of second kind. The associated 
hp can be obtained from eq. (\ref{eq63}) by replacing $\left( {t} \right)_{r} \to 
b_{r} (t)$.

\section{Burgers Equations and Heat Polinomials}
It is well known that the Hopf-Cole transformation \cite{10}
\begin{equation}
\label{eq66}
u(x,t) = {\frac{{\partial_x \Psi } }{{\Psi} }}{\rm ,}
\end{equation}
allows us to recast the heat equation (\ref{eq16}) as a non linear equation for the 
function $u(x,t)$, i.e.
\begin{equation}
\label{eq67}
\partial_t u =\partial_x^2 u + \partial_x (u^{2}) {\rm .}
\end{equation}
From the discussion of the previous section it follows that the hp are 
natural candidates for the solution of equation (\ref{eq67}) as
\begin{equation}
\label{eq68}
u(x,t) = {\frac{{nH_{n - 1} (x,t)}}{{H_{n} (x,t)}}}{\rm .}
\end{equation}

An analogous result can be obtained from the higher order heat equations 
(\ref{eq28}) for higher order Burgers equations \cite{10}. For example, in the case of 
eq. (\ref{eq28}) with $m = 3$, the transformation (\ref{eq66}) yields
\begin{equation}
\label{eq69}
\partial_t u = \partial_x (u^{3}) + 3(\partial_x u )^{2} + 3u \partial_x^2 u + \partial_x^3 u {\rm .}
\end{equation}
Eq. (\ref{eq69}) is satisfied by the higher order hp (hohp) of order 3, given by eq. 
(\ref{eq27}) with m=3.

It is obvious that any operator function $f(\hat {q})$, solution of the 
umbral heat equation can be used to get, via the corresponding Hopf-Cole 
transformation,
\begin{equation}
\label{eq70}
u = [f(\hat {q})]^{ - 1}\hat {p}f(\hat {q}){\rm .}
\end{equation}
 a solution of the Burgers equation (\ref{eq67})
\begin{equation}
\label{ub}
\partial_t \hat u = \hat p^2 \hat u + \hat p  ( \hat u^2 ){\rm .}
\end{equation}

We can use the case of the Laguerre 
derivative (\ref{eq44}, \ref{eq45})  to exemplify the problems one encounters 
when considering the nonlinear Burgers in umbral variables.
Taking advantage of the umbral correspondence and using the Hopf-Cole 
transformation
\begin{equation}
\label{eq71}
\omega (x,\partial _{t}^{ - 1} ) = \left( {\Psi (x,\partial _{t}^{ - 1} )} 
\right)^{ - 1}\Psi _{x} (x,\partial _{t}^{ - 1} ){\rm ,}
\end{equation}
we can write down the non-linear operator equation
\begin{equation}
\label{eq72}
\partial _{t} t\partial _{t} \omega (x,\partial _{t}^{ - 1} ) = \omega 
_{x,x} (x,\partial _{t}^{ - 1} ) + 2[\omega _{x} (x,\partial _{t}^{ - 1} 
)\omega (x,\partial _{t}^{ - 1} )]{\rm .}
\end{equation}
By projection onto a constant, we obtain from eq. (\ref{eq72}) a nonlinear 
functional equation.

The difficulty is now to interpret properly the umbral operator function 
$\omega (x,\partial _{t}^{ - 1} )$ and its functional form $\tilde {\omega 
}(x,t) = \omega (x,\partial _{t}^{ - 1} )\hat {1}$ in relation with the 
equation obtained by projecting eq. (\ref{eq72}). As will be explained in the 
following, the umbral non-linear equation (\ref{eq72}) should be interpreted as
\begin{equation}
\label{eq73}
 p_{t} \omega = \partial^2_{x} \omega + 2(\hat {\Omega} \omega ), \qquad
 \hat {\Omega}  = \partial _{x} \omega . 
\end{equation}
In other words eq. (\ref{eq73}) should be considered more like an identity rather 
than an equation.

The logical steps to check the validity of the above statement are the 
following:
\begin{enumerate}
\item We use the $\hat {T}$ transformation to derive the umbral form of the 
Burgers equation,
\item We use again $\hat {T}$ to infer the solution from that of the differential 
case.
\end{enumerate}
Let us consider a particular solution of the Laguerre heat equation (\ref{eq49}), 
the hybrid hp (\ref{eq50}) with $n = 2$:
\begin{equation}
\label{eq74}
H_{2,0} (x,\partial _{t}^{ - 1} ) = x^{2} + 2\partial _{t}^{ - 1} {\rm .}
\end{equation}
A solution of the Hopf-Cole transformation (\ref{eq71}) is given by
\begin{equation}
\label{eq75}
\omega (x,\partial _{t}^{ - 1} ) = 2(x^{2} + 2\partial _{t}^{ - 1} )^{ - 
1}x{\rm .}
\end{equation}
The solution (\ref{eq75}) in the coordinates $x,t$ is obtained by expanding its 
right hand side in power series. We find: 
\begin{equation}
\label{eq76}
\tilde {\omega}  = {\frac{{2}}{{x}}}{\sum\limits_{r = 0}^{\infty}  {\left( 
{{\frac{{ - 2\partial _{t}^{ - 1}} }{{x^{2}}}}} \right)}} ^{r}1{\rm .}
\end{equation}
Taking into account eq. (\ref{eq48}), eq. (\ref{eq76}) can be summed up and it reads:
\begin{equation}
\label{eq77}
\tilde {\omega}  = {\frac{{2}}{{x}}}e^{ - {\frac{{2t}}{{x^{2}}}}}{\rm .}
\end{equation}
The operator $\hat {\Omega}  = - 2{\frac{{x^{2} - 2\partial _{t}^{ - 1} 
}}{{(x^{2} + 2\partial _{t}^{ - 1} )^{2}}}}$ is clearly different  from the 
derivative with respect to $x$ of the function $\tilde {\omega} $.

The considerations developed so far suggest that, strictly speaking, the 
projection of eq. (\ref{eq73}) cannot be considered a non-linear equation but rather 
an umbral identity, satisfied by the ``umbralized'' operator solutions of 
the original equation. The umbral correspondences we have considered before 
are only relevant for linear equations. The effect of $\hat {T}$ on a 
non-linear equation does not present any difficulty provided we define the 
rules defining the transformation in a clear way and we properly understand 
the mathematical meaning of the obtained results.
According to eqs. (\ref{eq8}) we have
\begin{equation}
\label{eq78}
\hat {T}[f(x)^{2}]1 = f(\hat {q})\hat {T}f(x)1 = [f(\hat {q})]^{2}1 \ne 
[f(\hat {q})1]^{2}{\rm .}
\end{equation}
In the case of $\hat {q}$ given by eq. (\ref{eq11}) we find that
\begin{equation}
\label{eq79}
(\hat {q}1)^{2n} = H_{n} (x,y)^{2} \ne \hat {q}^{2n}1 = H_{2n} (x,y){\rm .}
\end{equation}
The remarks contained in eqs. (\ref{eq79}, \ref{eq78}) complete the considerations 
presented up above. Therefore, we can state the following result: given a non-linear differential equation for a scalar field 
$\Phi (x,t)$, its umbral image under the $\hat T$ map does not provide, by projection,a nonlinear scalar equation. The nonlinear umbral operator equation admits the solution
\begin{equation}
\label{eq80}
\Phi (\hat {q},t) = \hat {T}\Phi (x,t)\hat {T}^{ - 1}{\rm .}
\end{equation}
Following  this prescription, let us consider, as a final example,  the umbral counterpart of the sine-Gordon 
equation 
\begin{equation}
\label{eq81}
 {\left[ \partial_t^2 - \hat p^{2} \right]}\Phi (\hat {q},t)1 = \hat {F}\Phi (\hat {q},t)1, \qquad
 \hat {F} = {\sum\limits_{r = 1}^{\infty}  {{\frac{{\left( { - 1} 
\right)^{r}(2(\Phi (\hat {q},t))^{2r}}}{{(2r + 1)!}}}}} .
\end{equation}
$\Phi (\hat {q},t)1$ will not be the solution of eq. (\ref{eq81}). The 
equation
\begin{equation}
\label{eq82}
{\left[ \partial_t^2- \hat 
{p}^{2} \right]}\Phi (\hat {q},t)1 = [\sin (\Phi (\hat {q},t)1)],
\end{equation}
is not the umbral counterpart of the sine-Gordon equation.

\section{Concluding Remarks}

In this paper we have made extensive use of operator methods; in particular 
we have obtained the solutions of some differential equations using umbral 
calculus. For the ordinary heat equation the evolution operator coincides 
with a $\hat {T}$ transformation. In the more general example of eq. (\ref{eq39}) we 
have found that the solution depends on the operator
\begin{equation}
\label{eq83}
\hat {T}\hat {U}\hat {T}^{ - 1} = e^{(\hat {q} - \hat {p})t},
\end{equation}
where by $\hat {U} = e^{(x - \partial _{x} )t}$ we mean the evolution 
operator of the equation in ordinary space. The two equations (\ref{eq39}, \ref{eq40}) are formally equivalent as the $\hat T$  transformation leaves the Weyl group invariant.  

Problems may arise when we consider transformations which involve the 
time and in which the derivative operator is modified. In this case the 
exponential is not an eigenfunction of the corresponding derivative 
operator. For example, in the case of the Laguerre derivative we have
\begin{equation}
\nonumber
e^{\alpha \hat {t}}1 = e^{\alpha \partial _{t}^{ - 1}} 1 = {\sum\limits_{r = 
0}^{\infty}  {{\frac{{(\alpha t)^{r}}}{{r!^{2}}}}}} 1 = C_{0} (\alpha t) = 
I_{0} (2\sqrt {\alpha t} ){\rm ,}
\end{equation}
where $C_{0} (t)$ is the 0-th order Tricomi function and $I_{0} (2\sqrt 
{\alpha t} )$ the modified Bessel function of first kind \cite{12}. It is evident 
that the function $C_{0} (t)$ is an eigenfunction of the Laguerre derivative 
as\footnote{ The Laguerre derivative $\partial _{t} t\partial _{t} = 
\partial _{t} + t\partial^2 _{t} $ contains a  derivative of second order, therefore we 
may expect to have two eigenfunctions. However, the second eigenfunctions 
has a singularity at $t = 0$ and therefore it cannot be used to fulfil the 
condition $\hat {U}(0) = \hat {1}$.} 
\begin{equation}
\nonumber
(\partial _{t} t\partial _{t} )C_{0} (\alpha t) = \alpha C_{0} (\alpha 
t){\rm .}
\end{equation}
Accordingly we can write the solution of the equation
\begin{equation}
\nonumber
 {}_{L}D_{t} F(x,t)1 = xF(x,t)1 - \partial _{x} F(x,t)1, \quad
 F(x,0) = 1, 
\end{equation}
as
\begin{equation}
\nonumber
F(x,t) = e^{ - {\frac{{1}}{{2}}}\hat {t}^{2}}e^{x\hat {t}} = {\sum\limits_{r 
= 0}^{\infty}  {{\frac{{t^{r}}}{{\left( {r!} \right)^{2}}}}}} H_{n} (x, - 
{\frac{{1}}{{2}}}){\rm .}
\end{equation}
The above series expansion defines a 0-th order Bessel-Hermite function 
discussed in Refs. \cite{2,5}.

The above example shows that even in the case of linear equations, non 
trivial problems may arise. 

A further problem arises when discussing the meaning of the transformations from the physical point of view. The use of this transformations to solve a 
Schr\"{o}dinger equation in its umbral form may give rise to problems due to 
the non hermiticity of the variable involved. In fact such a property is not 
in general preserved by these transformation and this aspect of the problem 
requires a careful understanding which will be discussed in a forthcoming 
paper. 

In a forthcoming investigation we will discuss more in detail the 
mathematical properties of the evolution operator under umbral 
transformation and we will carefully treat the problems arising when time 
ordered products are involved \cite{13}. 

Work is also in progress for a better characterization of the solutions of 
the umbral Burgers equation, using perturbation methods.

\section*{Acknowledgments:}

D.L. wishes to thank P. Tempesta, P. Winternitz and O. Ragnisco for helpful 
discussions. D.L. was partially supported by PRIN Project ``SINTESI-2004'' 
of the Italian Minister for Education and Scientific Research and from the 
Projects \textit{Sistemi dinamici nonlineari discreti: simmetrie ed integrabilit\'{a}} and \textit{Simmetria e riduzione di equazioni differenziali di interesse fisico-matematico} of GNFM-INdAM. G.D. expresses his sincere appreciation to A. 
Siconolfi for a number of valuable discussions.

\end{document}